\documentclass[a4paper]{article}
\usepackage[latin1]{inputenc} 
\usepackage[T1]{fontenc} 
\usepackage{RR,RRthemes}
\usepackage{hyperref}
\pdfoutput=1

\usepackage[cmex10]{amsmath}
\usepackage{algorithmic}
\usepackage{algorithm}
\usepackage{verbatim}

\RRauthor{
Simplice DONFACK
\thanks{INRIA Saclay-Ile de France, Bat 490,
Universite Paris-Sud 11, 91405 Orsay France 
({\tt simplice.donfack@lri.fr}). 
} 
\and Laura GRIGORI
\thanks{INRIA Saclay-Ile de France, Bat 490, Universite Paris-Sud 11, 91405 Orsay France 
({\tt laura.grigori@inria.fr}). 
} 
\and William D. Gropp
\thanks{Department of Computer Science, University of Illinois at Urbana-Champaign, Urbana, IL 61801, USA 
({\tt wgropp@illinois.edu}). 
} 
\and Vivek Kale
\thanks{Department of Computer Science, University of Illinois at Urbana-Champaign, Urbana, IL 61801, USA 
({\tt vivek@illinois.edu}). 
} 
  
}

\RRtitle{Ordonnancement hybride statique/dynamique pour la factorisation deja optimisée des matrices denses}
\RRetitle{Hybrid static/dynamic scheduling for already optimized dense matrix factorization}

\RRabstract{We present the use of a hybrid static/dynamic scheduling strategy of the 
task dependency graph for direct methods used in dense numerical linear algebra. 
This strategy provides a balance of data locality, load balance, and
low dequeue overhead.
We show that the usage of this scheduling in communication avoiding
dense factorization leads to significant performance gains. On a 48
core AMD Opteron NUMA machine, our experiments show that we can
achieve up to 64\% improvement over a version of CALU that uses fully
dynamic scheduling, and up to 30\% improvement over the version of
CALU that uses fully static scheduling. On a 16-core Intel Xeon
machine, our hybrid static/dynamic scheduling approach is up to 8\%
faster than the version of CALU that uses a fully static scheduling or
fully dynamic scheduling. Our algorithm leads to 
speedups over the corresponding routines for computing LU  
factorization in well known libraries.  On the 48 core AMD NUMA machine, 
our best implementation is up to 110\% faster than MKL, while on the
16 core Intel Xeon machine, it is up to 82\% faster than MKL. Our 
approach also shows significant speedups compared with PLASMA on both
of these systems.
  }
  
\RRresume{
Nous présentons une nouvelle stratégie d'ordonnancement hybride statique/dynamique du
graphe de dépendance de tâches pour les méthodes directes utilisées en algèbre linéaire numérique dense.
Cette stratégie offre un équilibre entre la localité de données, l'équilibrage de la charge des processors  et la réduction de la charge de l'ordonnanceur de taches.
Nous montrons que l'utilisation de cette technique d'ordonnancement appliquée aux algorithmes de réduction de communications pour la factorisation des matrices denses conduit à des gains de performance significatifs. Sur une machine NUMA AMD opteron disposant de 48 cores, nos expériences montrent que nous pouvons atteindre des gains de performance de 64\% par rapport à la version de CALU qui utilise un ordonnancement statique, et jusqu'à 30\% par rapport à un ordonnancement dynamique. Sur une machine Intel Xeon disposant de 16 cores, notre approche est jusqu'à 8\% plus rapide que la version de CALU qui utilise un ordonnancement statique ou dynamique. Notre algorithme montre des améliorations importantes par rapport aux fonctions correspondantes à factorisation LU dans les libraires bien connus. Sur la machine AMD ayant 48 cores, la meilleur implémentation est jusqu'à 110\% plus rapide que MKL, tandis que sur la machine Intel xeon ayant 16 cores, elle est jusqu'à 82\% plus rapide que MKL. Notre approche montre aussi des accélérations significatives par rapport à PLASMA sur les deux machines.

 } 

\RRkeyword{dynamic scheduling, communication-avoiding, LU factorization} 

\RRmotcle{ordonnancement dynamique, reduction des communications, factorization LU} 
\RRprojet{Grand-large} 
\RRdomaine{1} 
\URFuturs 
\RCSaclay 

\begin{document}
\RRNo{0412}
\makeRR
\section{Introduction} 

One of the most important goals in high-performance computing is the 
design and development of efficient algorithms that can be portable for
a diverse set of node architectures and scale to emerging high-performance clusters 
with an increasing number of nodes. Many parallel scientific
applications are written using routines from 
numerical linear algebra libraries. Several
methods have been used to make such routines more tunable to 
a particular architecture, particularly due to the sensitivity 
of architectural parameters. 
In an effort to provide well-optimized BLAS~\cite{BLAS} that is
portable, numerical libraries such as GOTOBLAS~\cite{gotogotoblas} or ATLAS~\cite{ATLAS}
detect parameters of the user's system during installation, and tune
the library for a specific configuration. Cache-oblivious 
algorithms~\cite{frigo1999cache,toledo1997locality} avoid 
tuning for matrix computations by using the optimal data layout
independent of the size of the cache.



Despite various optimizations provided by vendors, the performance of
such routines may still be dramatically affected by architectural
characteristics that are hard to tune code for, particularly characteristics
that cause dynamic performance variations during execution of the routine.
In order to be scalable for future high-performance clusters(i.e. exascale), 
the code running within a node of a cluster must be tuned 
such that it achieves not simply ``high-performance'', but also
``performance consistency''~\cite{hoefler-noise-sim, vivek2010load}. 
Such static tuning techniques provide few guarantees on
performance consistency.  

\begin{figure}[htbp]
\label{fig:calu_static_profile}
 \centering
    \includegraphics[width=0.5\textwidth]{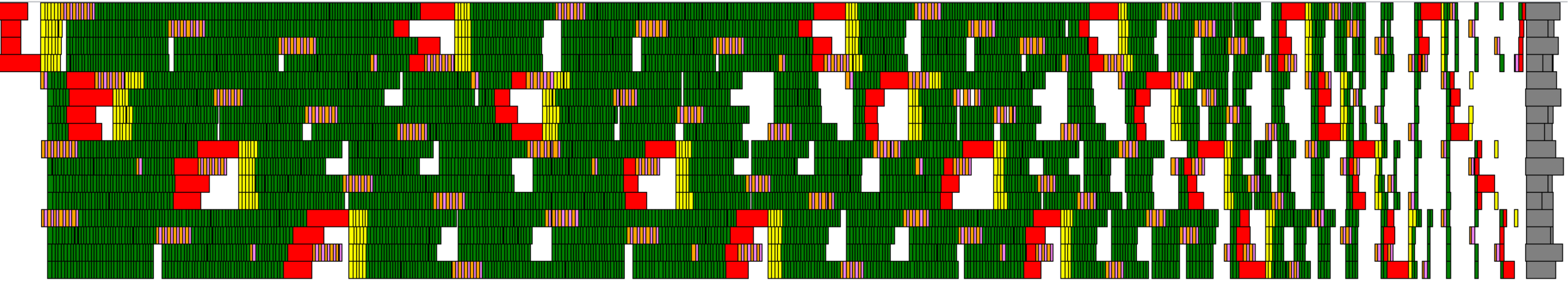}
    \caption{Profile of CALU using static scheduling on 16 cores of an
      AMD Opteron machine. }
\end{figure}

The problem becomes evident in a profile of a highly (statically
scheduled) optimized communication avoiding LU (CALU)
factorization. As can be seen in Figure~\ref{fig:calu_static_profile},
there are several pockets of thread idle time (shown through white
spaces), indicating that even a statically optimized and tuned code
still leads to idling cores during execution. This suggests that the
code is not able to completely harness the true (or peak) performance
of an architecture.  In addition, there are almost no patterns seen for the pockets
of idle time, suggesting a transient, dynamic performance variation of the
architecture that cannot necessarily be predicted through static techniques.

The emerging complexities of multi- and many-core architectures and
the need for performance consistency suggests making codes more
self-adaptive to the wide variety of different architectural
characteristics that are difficult to predict. Examples of such
self-adaptive techniques are
work-stealing~\cite{Blumofe99scheduling,blumofe1995cilk} and openMP
guided self-scheduling~\cite{guidedSched}.
 

 These self-adaptive strategies allow work to be \textit{dynamically
 scheduled} to cores (during execution of the linear algebra routine), rather than to be
\textit{statically scheduled} (before execution of the routine).
Using such dynamic scheduling techniques in numerical library routines
gives the advantage that available threads execute tasks as soon as
they are ready. Even if one thread becomes slow or inactive due to
transient performance variation induced by system events such
as I/O or OS daemons, the other threads will not be affected. 
The fundamental disadvantages of dynamic scheduling are 
the dequeue overhead and loss of data locality that
these strategies introduce; the conventional static scheduling
would not have caused overheads for such load balancing.
The dequeue overhead to pull a task from a
work queue can become non-negligible especially on an architecture with a
large number of cores. Dynamic scheduling provides no guarantee for threads to reuse data
resident in their local cache. The act of such dynamic migration of
data has a significant cost, especially on architectures with 
large differences in access time between cache and main memory. 
Thus, due to these potentially large scheduling overheads, 
 it may seem more appropriate to simply continue 
to use conventional static scheduling for these cases.

What is needed for such codes is a tunable scheduling strategy
that maintains load balance across cores while also maintaining data
locality and low dequeue overhead. To achieve this, we use a strategy that
combines static and dynamic scheduling. This approach was shown to be
successful on regular mesh computations~\cite{vivek2010load}.
 This tunable scheduling strategy allows us
to flexibly control the percentage of tasks that can be scheduled
dynamically; this gives to a knob to control load balancing so that
it occurs only at the point in computation when the benefits it
provides outweighs the costs it induces. 
On NUMA machines where remote memory access is costly,
the percentage of work scheduled dynamically should be small enough to avoid 
excessive cache misses, but large enough to keep the cores busy during idle times in the static part.



In this work, we show the effectiveness of this method in the context
of already highly-optimized dense matrix factorizations. 
We focus in particular on communication avoiding LU (CALU), a recently
introduced algorithm which offers minimal
communication~\cite{grigori2008communication}.  We choose LU because
of its tight synchronization and communication constraints.  Most of
the discussion applies to other routines in dense numerical linear
algebra also.  Our prior work on multi-threaded
CALU~\cite{donfack2010adapting} was based on dynamic scheduling.
The algorithm performed well on tall and skinny matrices, but became
less scalable on square matrices with increasing numbers of
processors. When the input matrix is tall and skinny, the fast
factorization of the panel usually surpasses the various types of
scheduler optimizations. However, when the matrix is large, it becomes important
to modify the properties of the scheduler to take into account memory
bandwidth bottlenecks and data locality.

In our hybrid scheduling approach, the percentage of computation
that allotted to be dynamic scheduled can be tuned based on the underlying
architecture and the input matrix size.  An important factor that
impacts performance is the data layout of the input matrix. Hence, we
also investigate three data layouts: a classic column major format, a
block cyclic layout, and a two level block layout.  Table
\ref{tab:design_space} describes the design space we explore.
By avoiding load balancing until it is absolutely needed, we are able
to acheive significantly higher performance gains over a fully static or a fully dynamic scheduling strategy, 
and also can provide better performance compared to two well-known
numerical linear algebra libraries, MKL and PLASMA.  While the paper
focuses specifically on CALU, most of the discussion applies to other
methods of factorization such as QR, rank-revealing QR, and to a certain
degree the Cholesky and $LDL^T$ factorizations.






\begin{table}[htbp]
	\begin{tabular}{|l|l|l|l|}
	\hline							
				 & \multicolumn{3}{|c|}{CALU}\\
	\hline
	Data Layout $\backslash$ Scheduling & Static 					& Dynamic 	&static (number\%\\
																			&  								&						&dynamic)\\
	\hline
	Block cyclic layout (BCL) 					& $\surd$ 				& $\surd$		&$\surd$\\
	\hline
	2-level block layout (2l-BL)				& $\surd$ 				&$\surd$		&$\surd$\\
	\hline
	Column major layout (CM)  											&  								&	$\surd$ 	&	\\

	\hline
	\end{tabular}
	\caption{Design space. In the hybrid version, number\% represents the percentage of the dynamic part}
	\label{tab:design_space}
\end{table}
This paper is organized as follows. In the section 2, we briefly
give a background of LU and CALU factorization. In the section 3, we
show how to combine static and dynamic scheduling to achieve good
performance in numerical librairies. In section 4, 
we discuss the data layout we use for the matrices. In section 5, we present
experimental results. In section 6, we present a theoretical analysis. 
In section 7, we present a discussion and broader impact to future machines(e.g. exascale). 
In section 8, we discuss relevant related work. In section 9, we conclude the paper and discuss
future work.

\section{Direct methods in dense linear algebra}
\label{sec:calu}

In this section we briefly introduce the direct methods of
factorization, and in particular the LU factorization and its
communication avoiding variant, CALU.  The LU factorization decomposes
the input matrix $A$ of size $m \times n$ into the product of $L \cdot
U$, where $L$ is a lower triangular matrix and $U$ is an upper
triangular matrix.  In a block algorithm, the factorization is
computed by iterating over blocks of columns (panels).  At each
iteration, the LU factorization of the current panel is computed, the
$U$ factor of the current block row is determined, and then the
trailing matrix is updated.  This last step is computationally the
most expensive part of the algorithm.  It can be performed efficiently
since it exploits BLAS 3 operations, and it exposes parallelism.  The
panel factorization, even if it does not dominate the computation in
terms of flops, is a bottleneck in terms of parallelism and
communication.  This is because the update of the trailing matrix can
be performed only once the panel is factored.  Hence, it is important
to perform the panel factorization as fast as possible.  For example,
the multithreaded LAPACK~\cite{LAPACK} performs the
panel factorization sequentially, and this leads to poor performance, even if the
update is performed in parallel.

However, due to partial pivoting, its parallelization is not an easy
task.  The panel factorization requires communication (or
synchronization in a shared memory environment) for the computation of
each column, and this leads to an algorithm which communicates
asymptotically more than what the lower bounds on communication
require.  An efficient sequential algorithm is the recursive LU
factorization \cite{toledo1997locality, gustavson97:_recur}.  However,
a parallelization of this approach will likely have scalabiltity
limits due to the recursive formulation of the algorithm.

Communication avoiding algorithms introduced in the last years provide
a solution to this problem.  In the case of the LU factorization, its
communication avoiding version CALU~\cite{grigori2008communication}
uses a different pivoting strategy, tournament pivoting, which is
shown to be as stable as partial pivoting in practice.  With this
strategy, the panel factorization can be efficiently parallelized, and
the overall algorithm is shown to provably minimize communication.
The panel factorization, refered to as TSLU, is computed in two steps.
The first preprocessing step identifies, with low communication cost,
pivots that can be used for the factorization of the entire panel.
These pivots are permuted into the diagonal positions, and the second
step computes the LU factorization with no pivoting of the entire
panel.  The preprocessing step is performed as a reduction operation,
with LU factorization with partial pivoting being the operator used at
each step of the reduction.  We use a binary tree for the reduction,
which is known to minimize communication.

In CALU, the panel factorization remains on the critical path.
However, current research indicates that this is required for
obtaining a stable pivoting strategy and a stable factorization.  A
different pivoting strategy known as block pairwise pivoting removes
the panel factorization from the critical path, but this strategy
requires more investigation in terms of stability.  This approach was
explored in previous versions of PLASMA~\cite{buttari07} and
FLAME~\cite{g.08:_progr}.

The scheduling strategy that we present in the following section
relies on the task dependency graph of CALU.  We consider that the
input matrix $A$ is partitioned into blocks of size $b \times b$ as
\begin{eqnarray*}
\label{eq:Apart}
A =
  \begin{pmatrix}
    A_{11} & A_{21} & \ldots & A_{1N} \\
    A_{21}  & A_{22} & \ldots & A_{2N} \\
    \vdots & \vdots &  & \vdots \\
    A_{M 1} & A_{M 2} & \ldots & A_{M N}
  \end{pmatrix},
\end{eqnarray*}
where $M = m/b$ and $N = n/b$.  

The task dependency graph is obtained by considering that the
computation of each block $A_{ij}$ is associated with a task.  We
distinguish the following tasks:
\begin{itemize}
\item \textbf{task P}: participates in the preprocessing step of the
  panel factorization TSLU.
\item \textbf{task L}: computes part of the $L$ factor of the current
  panel, by using the pivots identified in task P.
\item \textbf{task U}: computes a block of the $U$ factor in the
  current row.
\item \textbf{task S}: updates a block of the trailing matrix.
\end{itemize}

\begin{figure}[htbp]
  \begin{center}        
    \includegraphics[angle=0,scale=0.35]{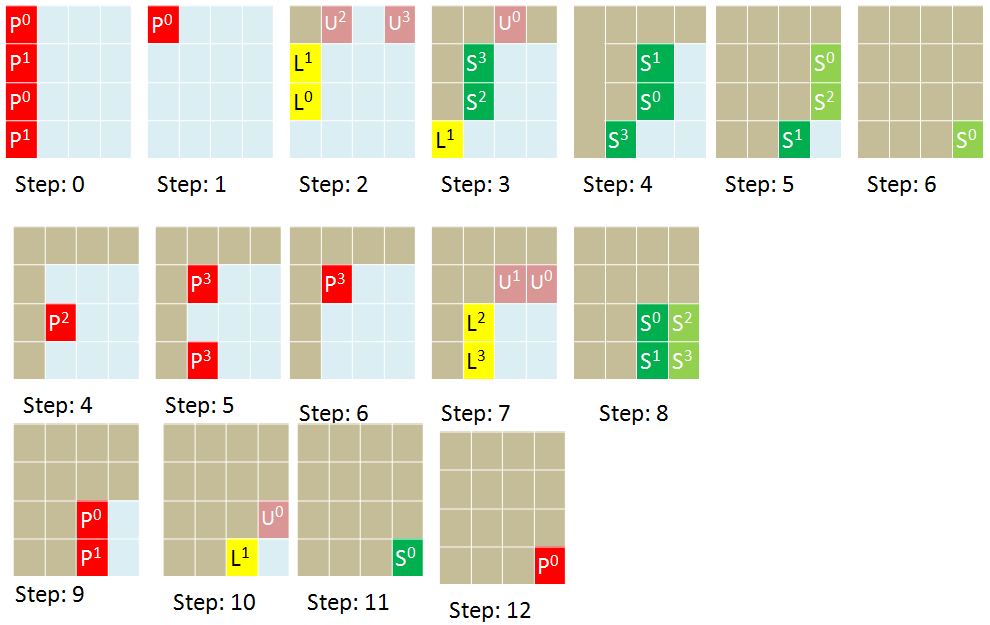}
    \caption{Example of execution of CALU static(20\%) dynamic on a
      matrix partitioned into $4 \times 4$ blocks using P=$4$ threads.}
    \label{fig:matrixExecution}
  \end{center}
\end{figure}

\begin{figure}[htbp]
  \begin{center}        
    \includegraphics[angle=0,scale=0.35]{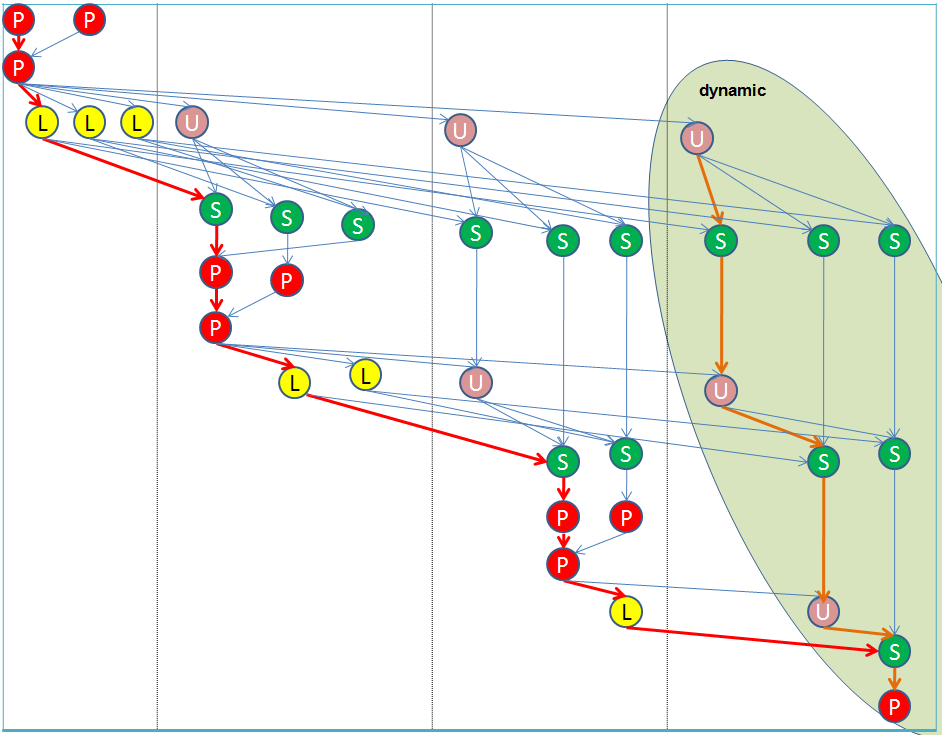}
    \caption{Task dependency graph of CALU static/dynamic of a matrix
      partitioned into $4 \times 4$ blocks. The red arrows indicate
      the critical path of the static section of the algorithm, while
      the green arrows indicate the critical path of the dynamic
      section of the algorithm. }
    \label{fig:dag_path}
  \end{center}
\end{figure}

With these notations, a matrix partitioned into $4 \times 4$ blocks is
showed in Figure \ref{fig:matrixExecution}, and its task dependency
graph (DAG) is displayed in Figure~\ref{fig:dag_path}.

\section{Scheduling based on a hybrid static/dynamic strategy}

In this section, we describe our hybrid static/dynamic scheduling
strategy that aims at exploiting data locality, ensuring good load
balance, reducing scheduling overhead, and being able to adapt to
dynamic changes in the system.  The hybrid scheduling is obtained by
spliting the task dependency graph into two parts, a first part which
is scheduled statically, and a second part which is scheduled
dynamically.  Given a parameter $N_{static}$, tasks that operate on
blocks belonging to the first $N_{static}$ panels are scheduled
statically, while tasks that operate on blocks belonging to the last
$(N - N_{static})$ panels are scheduled dynamically.  Hence, the tasks
that lie on the critical path of the algorithm are scheduled
statically.  During the factorization, each thread executes in
priority tasks from the static part, to ensure progress in the
critical path of the algorithm.  When there are no ready static tasks,
then the thread picks up a task from the dynamic part.  Thus, the two
parts of the task dependency graph are not independent. 

Algorithm~\ref{alg:CALU_static_dynamic} describes CALU with hybrid
static/dynamic scheduling.  In the static part of the DAG, the matrix
is distributed to threads using a classic two-dimensional block-cyclic
distribution.  The algorithm proceeds as follows.  For the first
$N_{static}$ iterations, once the panel $K$ is factored statically,
several tasks become ready.  These tasks are grouped into two distinct
sets.  The first set is formed by tasks that update blocks $A_{:,J}$,
with $K+1 \leq J \leq N_{static}$. These tasks are scheduled
statically.  They are inserted into the queue of ready tasks of the
thread which owns the blocks $A_{:,J}$.  The second set is formed by
tasks that operate on blocks $A_{:,J}$ with $N_{static} < J \leq N$.
These tasks are scheduled dynamically, they are inserted in a shared
global queue of ready tasks.  For clarity of the presentation, the
algorithm does not present the insertion of ready tasks in the dynamic
queue. 

For the last $N - N_{static}$ iterations, the algorithm uses a fully
dynamic scheduler. While the same pattern of execution is used as in
the static part, the main difference is that now the tasks are
scheduled dynamically when they are ready. 

The static part of Algorithm~\ref{alg:CALU_static_dynamic} uses the
routine "dynamic\_tasks()", which is described in Algorithm
\ref{alg:dynamic_task}.  This routine selects one task in the dynamic
part of the matrix when a thread requests it.  The task is selected by
traversing the DAG associated with the dynamic part using a depth-
first search approach.  With this approach, the columns are updated
from left to right.  This ensures that the execution follows in
priority the critical path when the algorithm will reach the dynamic
section.  The tasks selected by the routine are removed from the
global queue.

We note that a static/dynamic approach makes two critical paths
appear.  The first path corresponds to the critical part of the task
dependency subgraph scheduled statically.  In our case, this
corresponds to the critical path of the whole task dependency graph of
CALU.  The second path corresponds to the critical path of the task
dependency subgraph scheduled dynamically.  The two paths are
displayed in Figure~\ref{fig:dag_path} on our task dependency graph
example.  The second path is executed in parallel with the first
one. We consider the second path as important as the first one;
otherwise, the algorithm can stagnate when it arrives at the dynamic
section.  

We use the following notation and routines in
Algorithm~\ref{alg:CALU_static_dynamic}:
\begin{itemize}
\item \textit{I\_Own(panel(K))}: returns true if the thread executing
  this instruction owns a block of panel $K$.
\item \textit{I\_Own(block-row(K))}: returns true if the thread
  executing the instruction owns a block of block-row $K$.
\item \textbf{task P}: each thread executing this task performs a
  reduction operation to identify $b$ pivots that will be used for the
  panel factorization of the current panel.  The reduction operator is
  Gaussian elimination with partial pivoting, and for this the best
  available sequential algorithm can be used.  In our experiments we
  use recursive LU~\cite{toledo1997locality}.
\item "\textbf{do task L} (on $L_{K:M,K}$) \textbf{in parallel}":
  computes the $L$ factor of panel $K$ in parallel.  In the static
  section, each thread computes the blocks of $L$ of panel $K$ that it
  owns.  In the dynamic section, available threads compute blocks of
  $L$ of panel $K$ until the whole panel is finished.
\item "\textbf{do task S} (on $A_{K+1:M,J}$) \textbf{in parallel}":
  panel $J$ is updated in parallel. In the static section, each thread
  updates the blocks of panel $J$ that it owns (if any).  In the
  dynamic section, available threads update blocks of panel $J$ until
  the whole panel is finished. 
\end{itemize}

\begin{algorithm}[htpb]
\caption{CALU with hybrid static/dynamic scheduling}
\label{alg:CALU_static_dynamic}
{\small
\begin{algorithmic}[1]
\STATE{\textbf{Input:} $m \times n$ matrix $A$, block size $b$,
  percentage of dynamic section $d_{ratio}$}
\STATE{$N_{static} = N*(1-d_{ratio})$}

\STATE /* static section*/
\FOR{$K$ = $1$ to $N_{static}$}
\STATE \quad \textbf{if} I\_Own(panel(K)) \textbf{then}
\STATE \quad \quad \textbf{do task P} Preprocessing of TSLU($A_{K:M,K}$) \textbf{in parallel}			 
\STATE \quad \textbf{endif}
\STATE \quad \textbf{while} $task P[K]$ not done 
\STATE \quad \quad \textbf{do} dynamic\_tasks()
\STATE \quad \textbf{end while}
\STATE \quad Let $\Pi_{KK}$ be the permutation performed for panel $K$ such that $\Pi_{KK}A_{1:b, 1:b} = L_{KK}U_{KK}$
\STATE \quad \textbf{do task L} $L_{K:M, K} = A_{K:M, K} U_{KK}^{-1}$ \textbf{in parallel}	
\STATE \quad \textbf{if} I\_Own(block-row(K)) \textbf{then}
\STATE \quad \quad \textbf{for} $J$ = $K+1$ to $N_{static}$
\STATE \quad \quad\quad \textbf{if} I\_Own(panel(J)) \textbf{then}
\STATE \quad \quad\quad\quad \textbf{do right swap} $A_{K:M, J} = \Pi_{KK} A_{K:M, J}$
\STATE \quad \quad\quad\quad \textbf{do task U} $U_{K,J} = L_{KK}^{-1} A_{K,J} $ 
\STATE \quad \quad\quad \textbf{endif}
\STATE \quad \quad \textbf{end for}
\STATE \quad \textbf{endif}		
\STATE \quad \textbf{for} $J$ = $K+1$ to $N_{static}$
\STATE \quad \quad \textbf{if} I\_Own(panel(J)) \textbf{then}
\STATE \quad \quad\quad \textbf{while} the block $U_{K,J}$ not computed 
\STATE \quad \quad\quad\quad \textbf{do} dynamic\_tasks()
\STATE \quad \quad\quad \textbf{end while}
\STATE \quad \quad\quad \textbf{do task S} $A_{K+1:M,J} -= L_{K+1:M,K} U_{K,J}$  \textbf{in parallel} 			 
\STATE \quad \quad \textbf{endif}
\STATE \quad \textbf{end for}
\ENDFOR
\STATE /* dynamic section*/
\FOR{$K$ = $N_{static}$ to $N$}
\STATE \quad \textbf{do task P} Preprocessing of TSLU($A_{K:M,K}$)
\textbf{in parallel and dynamically}			 
\STATE \quad Let $\Pi_{KK}$ be the permutation performed for panel $K$ such that $\Pi_{KK}A_{1:b, 1:b} = L_{KK}U_{KK}$
\STATE \quad \textbf{do task L} $L_{K:M, K} = A_{K:M, K} U_{KK}^{-1}$
\textbf{in parallel and dynamically}	
\STATE \quad \textbf{for} $J$ = $K+1$ to $N$ \textbf{do in parallel
  and dynamically}
\STATE \quad \quad \textbf{do task U} $A_{K,J} = L_{KK}^{-1} A_{K,J} $
\STATE \quad \textbf{end for}	
\STATE \quad \textbf{for} $J$ = $K+1$ to $N$ \textbf{do in parallel
  and dynamically}
\STATE \quad \quad \textbf{do task S} $A_{K+1:M,J}$ -= $L_{K+1:M,K} U_{K,J}$ \textbf{in parallel} 			 
\STATE \quad \textbf{end for}
\ENDFOR
\STATE /* Apply permutations to the left */
\STATE  $L_{1:M,1:N} = \Pi_{NN} \ldots \Pi_{11} L_{1:M,1:N}$  \textbf{in parallel} \quad\quad/* dlaswap */
\end{algorithmic}
}
\end{algorithm}

\begin{algorithm}[htpb]
\caption{dynamic\_task}
\label{alg:dynamic_task}
\begin{algorithmic}[1]
\STATE Let $K_0$ be the panel currently computed in the static
section.
\FOR{$J$ = $N\_STATIC+1$ to $N$}
\STATE \quad \textbf{for} $K$ = $1$ to $K_0-1$ \textbf{do}
\STATE \quad \quad \textbf{if} $U_{K,J}$ not computed
\STATE \quad \quad \quad \textbf{do task U} $U_{K,J} = L_{KK}^{-1}
A_{K,J} $ 
\STATE \quad \quad \textbf{endif}
\STATE \quad \quad \textbf{if} $A_{:,J}$ not updated by panel $K$
\STATE \quad \quad \quad \textbf{do task S} $A_{I,J} = A_{I,J} -
L_{I,K}*U_{K,J}$ for all $I$ with $K+1 \leq I \leq M$
\STATE \quad \quad \textbf{endif}
\STATE \quad \textbf{endfor}
\ENDFOR
\end{algorithmic}
\end{algorithm}

We show an example of execution of our algorithm on the same matrix
formed by $4 \times 4$ blocks from Figure \ref{fig:matrixExecution}
and its task dependancy graph from Figure \ref{fig:dag_path}.  In
Figure \ref{fig:matrixExecution}, the exponent indicates the thread
which executes the task. At steps 5 and 6, we observe that instead of
becoming idle while waiting for the completion of the factorization of
the third panel, two threads execute a task from the dynamic section.
This avoids unnecessary idle time.

In our current work, the percentage of the dynamic part $d_{ratio}$ is
a tuning parameter, which determines the number of panels that will be
executed statically ($N_{static}$) or dynamically ($N- N_{static}$).  It
is therefore possible to switch from a 100\% static version to a 100\%
dynamic version.  In practice, a particular scheduling technique can be highly
efficient on one architecture, but less efficient on another. In the
experiemental section, we show that the best scheduling strategy
depends not only on the architecture on which it is executed, but also on
the size of the matrix, the data layout of the matrix, and the number
of processors.  The flexibility to choose the percentage of the
dynamic section in our algorithm will allow it to adapt on different
architectures.

\begin{figure}[htbp]
  \begin{center}        
    \includegraphics[angle=0,scale=0.4]{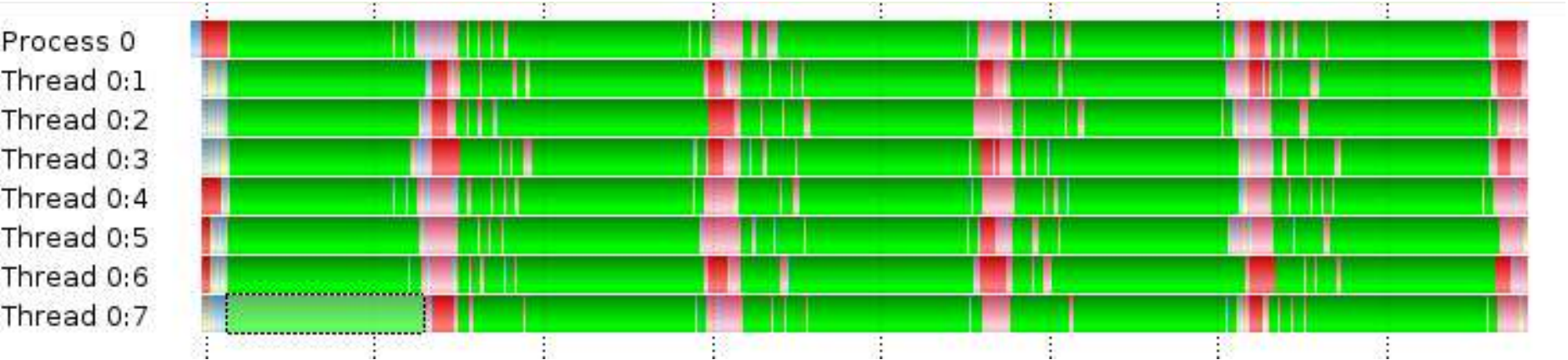}
    \caption{First steps of the factorization of a matrix of size
      $5000 \times 5000$ using a static (20\% dynamic) scheduling} 
    \label{fig:fig_overlappingPanel}
  \end{center}
\end{figure}

As explained in section~\ref{sec:calu}, for factorizations that use
some form of pivoting as CALU or Gaussian elimination with partial
pivoting, the panel factorization lies on the critical path. In a
fully static approach, this can be a bottleneck and may cause
inactivity due to lack of tasks.  The combined static/dynamic
scheduling helps to overcome this problem.  Threads that are idle
waiting for the completion of the panel factorization perform tasks
from the dynamic part.  This is illustrated in
Figure~\ref{fig:fig_overlappingPanel} where a static (20\% dynamic)
scheduling is used to factor a matrix of size $5000 \times 5000$.  The
red tasks represent the panel factorizations and the green tasks
represent the updating computation.  We observe that some of the
threads finish earlier than others the panel factorization.  In a
fully static approach, they would become idle.  In the hybrid
approach, they execute tasks from the dynamic section, and there is
almost no idle time in this example.

Several other optimizations are used in our algorithm that are
important for its performance, but we do not describe them in
detail in this paper.  For example, the static section employs
look-ahead, a technique used in dense factorizations to allow the
panel factorizations to be performed as quickly as possible.  The
granularity of the tasks used during the update of the trailing matrix
has a direct impact on the performance of BLAS 3 operations (dgemm or
cgemm in our case).  The best granularity is a trade-off between
parallelism (there should be enough tasks to schedule) and BLAS 3
performance.  In the static section, a thread can update the blocks it
owns one by one, or it can group them together and update using one
single call to BLAS 3.  The latter option leads to a better
performance of BLAS 3, and also to a reduction in the number of
messages transfered (if an appropriate data layout is used).  However
the number of words transfered stays the same.  In our experiments,
the threads update the trailing matrix by using blocks of size $kb, k
\geq 1$, with $k = 3$.

\section{Data Layout}
\label{sec:data_layout}

Classic libraries such as Lapack and Scalapack store the matrices using a
column major layout.  However, novel algorithms that minimize
communication such as CALU require the usage of novel data layouts, based
on blocking or recursive blocking.  In this paper, we investigate the
impact on performance of two data layouts that are adapted to our
algorithm.  We describe them in the following.

\subsection{Block cyclic layout (BCL)}

This layout aims at enabling data locality in the static section of
our algorithm.  The static section considers that the matrix is
distributed using a 2D block cyclic layout over a 2D grid of threads.
Then, during the algorithm, each thread modifies the blocks that it
owns. The block cyclic layout stores contigously in memory, for each
thread, the blocks that it owns.  In other words, the matrix is
partitioned into as many submatrices as threads. Each submatrix is
stored in memory using a column major layout.  Note that a submatrix
is formed by blocks issued from the 2D block cyclic layout; that is,
their column indices and row indices are not contiguous (except inside
the small blocks of size $b \times b$).

The block cyclic layout is displayed at the left of figure
\ref{fig:matrixLayout}.  The matrix was partitioned into blocks of
size $2 \times 2$ which were distributed using the 2D block cyclic
layout over a grid of 4 threads.  The main avantage of this storage
compared to full column major layout as implemented in LAPACK is that
in the static section, the data of each thread is stored contiguously
in its local memory.
Another avantage is the possibility of improving BLAS 3 performance,
described in the previous section.  Each thread can simply call dgemm
(or cgemm) on a block which can be larger than $b \times b$.

\subsection{Two level block layout (2l-BL)} 

\begin{figure}[htbp]
  \begin{center}        
    \includegraphics[angle=0,scale=0.45]{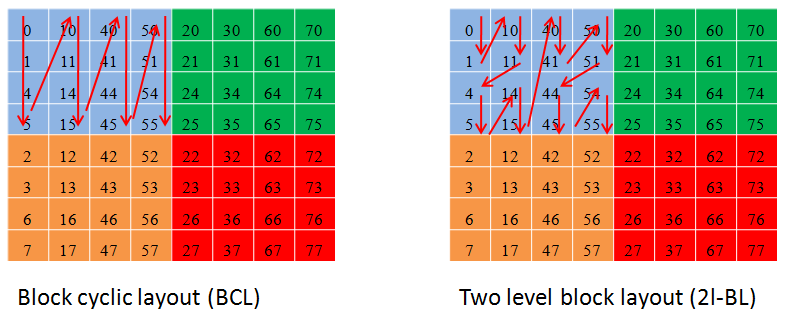}
    \caption{Data layout.  The figure at left displays the
      partitioning of the matrix into four blocks using a 2D block
      cyclic layout (BCL) based on blocks of size $b \times b$.  Each
      of the four blocks is stored contiguously in memory.  The figure
      at right displays the two level block layout (2l-BL) layout,
      which further stores contigously in memory blocks of size $b
      \times b$ for each of the four blocks.}
    \label{fig:matrixLayout}
  \end{center}
\end{figure}

This layout can be seen as a recursive block layout, with the recursion
being stopped at depth two (with the exception that at the first
level, the matrix is partitioned using a block cyclic layout).  At the
second level of the recursion, the submatrix belonging to each thread
is further partitioned into blocks of size $b \times b$ and each block
is stored contigously in memory.  This partitioning is shown at the
right of Figure \ref{fig:matrixLayout}.  The main avantages of this
storage is that, with an appropriate value of $b$, the block
(sometimes refered as tile) can fit in cache at some level of the
hierarchy.  Hence any operation on the block can be performed with no
extra memory transfer.

However, with this layout, it is not straightforward to increase the
size of the blocks used during the update of the trailing matrix using
BLAS 3.  This would require a copy of the data, which could add extra
time.  We do not explore this option in this paper.

\section{Experimental results}

In this section we evaluate the performance of our algorithms on a
four-socket, quad-core machine based on Intel Xeon EMT64 processor and
on an eight-socket, six-core machine based on AMD Opteron processor
running on Linux. The Intel machine has a theoretical peak performance
of 85.3 Gflops/second in double precision. Each core has a frequency
of 2.67GHz, a private L1 cache of size 32 Kbytes, an L2 cache of size
512 Kbytes, and an L3 cache of size 8192 Kbytes shared with the others
cores. The AMD machine has a theoretical peak performance of 539.5
Gflops/second in double precision.  Each core has a frequency of 2.1
GHz, a private L1 cache of size 64 Kbytes, a private L2 cache of size
512 Kbytes, and an L3 cache of size 5118 Kbytes shared with the other
cores of the same socket.

We first present the performance of the hybrid static/dynamic
scheduling compared to a fully static and a fully dynamic scheduling,
while also discussing the impact of the data layout on performance.
We then compare the performance with the corresponding routines from
MKL 10.3.2 vendor library and PLASMA 2.3.1.

\subsection{Performance of static/dynamic scheduling}

In the following, \textit{CALU static} refers to the version of CALU
based on fully static scheduling, while \textit{CALU dynamic} refers
to the version of CALU based on fully dynamic scheduling. CALU
static/dynamic refers to the version of CALU based on combined static
and dynamic scheduling.  When we want to identify the percentage of
the dynamic part, we use CALU static(\textit{number\%} dynamic), where
\textit{number\%} specifies the percentage of the computation
scheduled dynamically.

\subsubsection{Comparison with static and dynamic scheduling using block cyclic layout}

We first discuss the performance of CALU using a block cyclic layout.
As explained earlier, one of the advantages of this layout is that
during the update of the trailing matrix, we can call dgemm on larger
blocks by grouping together blocks that are stored in the same memory.
While we can group together blocks that share the same rows or the
same columns, we choose the latter option such that the algorithm can
make progress on its critical path.

\begin{figure}[htbp]
  \begin{center} \includegraphics[scale=0.3]{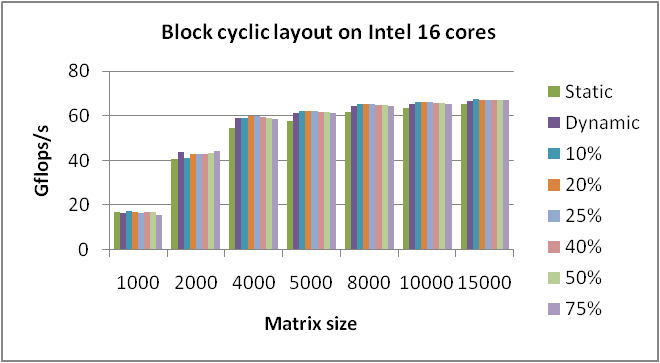} \caption{Performance
    of CALU with static/dynamic scheduling on Intel 16-core
    machine. The percentage of the dynamic part is varied from 10\% to
    75\%. The matrix of size $M=N=5000$ is stored using block cyclic
    layout.} \label{fig:bcl_intel_all_16cores} \end{center}
\end{figure}

Figure \ref{fig:bcl_intel_all_16cores} shows the performance of CALU
static, CALU dynamic and CALU static/dynamic with varying the
percentage of the dynamic scheduled work on the 16 core Intel Xeon machine.  We
observe that hybrid static/dynamic scheduling is more efficient than
either of static or dynamic scheduling.  In particular, CALU static(10\%
dynamic) is 8.20\% faster than CALU static, and is 1.4\% faster than CALU
dynamic. However,the difference obtained by varying the percentage of the
dynamic section is not significant.
On this machine, the static scheduling is the least efficient, while
the dynamic scheduling is closer to the best performance obtained by
the static/dynamic approach.  This performance is explained by the
fact that the scheduler overhead and the cache miss penalties of the
dynamic approach are less significant than the idle time introduced by load
imbalance in the static approach.

\begin{figure}[htbp]
  \begin{center}
    \includegraphics[scale=0.3]{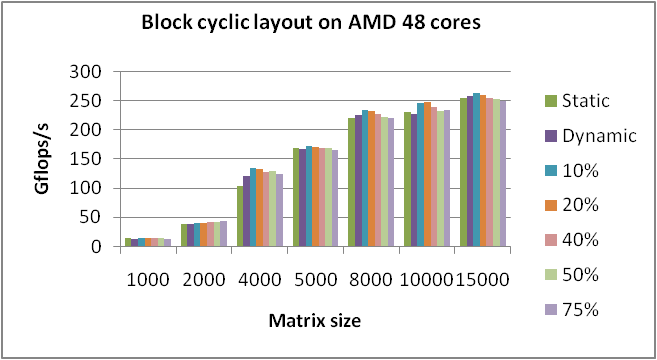}
    \caption{Performance of CALU with static/dynamic scheduling on the
    48 core AMD opteron machine. The percentage of the dynamic part is varied from 10\% to
    75\%. The matrix is stored using the block cyclic layout.}
    \label{fig:bcl_amd_all_16cores} \end{center}

\end{figure}

Figure \ref{fig:bcl_amd_all_16cores} shows the performance obtained on
the 48 core AMD Opteron machine. On NUMA machines, the memory latency
plays an important role on performance.  Static scheduling is very
appropriate in this case because of its good use of data
locality. However, the best performance is obtained by combining
static with a small percentage of dynamic (10\% or 20\%), which is
sufficient to reduce the thread idle time that cannot be handled
by using purely static scheduling.

\begin{figure}
   \begin{minipage}[c]{.49\linewidth}
      \includegraphics[height=0.65\linewidth, width=\linewidth]{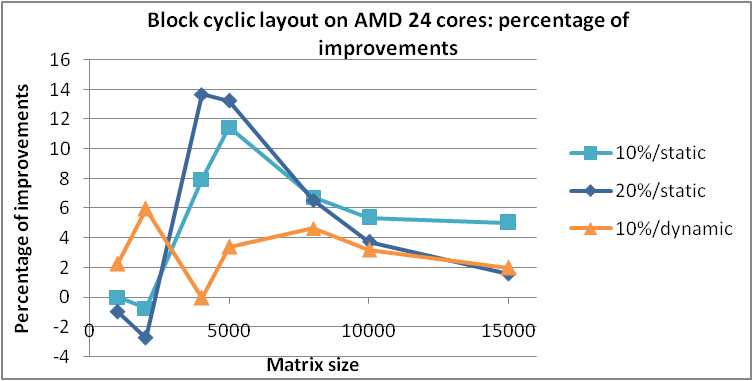} 
      a. Experiments on 24 cores
   \end{minipage} \hfill
   \begin{minipage}[c]{.49\linewidth}
      \includegraphics[height=0.65\linewidth, width=\linewidth]{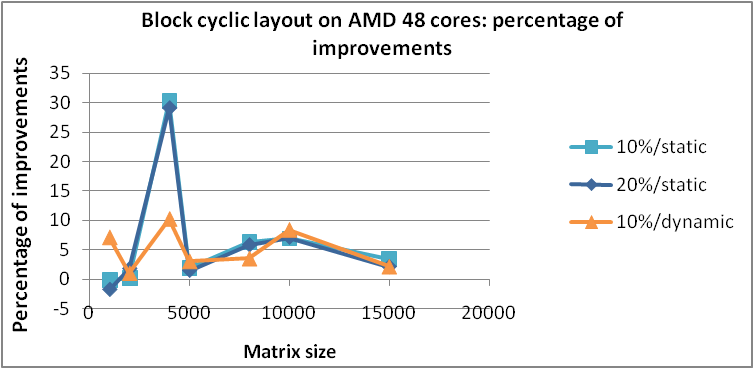}
      b. Experiments on 48 cores
   \end{minipage}
   \caption{Percentage of improvement of CALU static(10\% dynamic) and
   CALU static(20\% dynamic) over CALU static and CALU dynamic on the
   AMD  48-core machine. The matrix is stored using a block cyclic
   layout.}

   \label{fig:bcl_precentage_improvements}
\end{figure}

Figure \ref{fig:bcl_precentage_improvements} shows the percentage of
improvement of CALU static(10\% dynamic) and CALU static(20\% dynamic)
over CALU static and CALU dynamic.  The best improvement is observed
on 48 cores with $M=N=4000$, where CALU static(10\% dynamic) is 30.3\%
faster than CALU static and 10.2\% faster than CALU dynamic. For
$M=N=10000$ on 48 cores, CALU static(10\% dynamic) is 6.9\% faster
than CALU static and 8.4\% faster than CALU dynamic. 
This suggests that, especially for smaller matrices, using just a small percentage of dynamic 
scheduling can provide significant performance benefits.  
When we do these experiments using only 24 cores, CALU
static(20\% dynamic) is slightly faster than CALU static(10\% dynamic). This
suggests that in some cases, increasing the percentage of dynamic scheduling could
lead to better performance, and that this percentage should be appropriately tuned. 


\subsubsection{Comparison with static and dynamic scheduling using 2-level block layout}
\begin{figure}[htbp]
  \begin{center}        
    \includegraphics[scale=0.3]{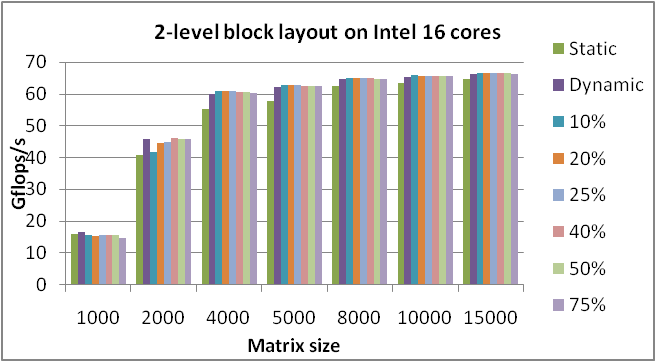}
    \caption{Performance of CALU with static/dynamic scheduling on
    Intel 16 core machine. The percentage of the dynamic part is
    varied from 10\% to 75\%. The matrix is stored using 2-level block
    layout.} 

    \label{fig:2lbl_intel_all_16cores}
  \end{center}
\end{figure}

We now discuss the performance of CALU when using a 2-level block
layout.  Figure \ref{fig:2lbl_intel_all_16cores} shows the performance
obtained on 16 cores Intel Xeon machine.  The behavior is the same as
observed with the block cyclic layout.  Increasing the percentage of
dynamic in CALU static/dynamic does not have an important impact on
performance.  Again, the static scheduling is less efficient than all
the other approaches.  The best improvement is obtained with CALU
static(10\% dynamic) for $M=N=4000$, when it is 10.6\% faster than
static and 1.7\% faster than dynamic. 
  
\begin{figure}[htbp]
  \begin{center}        
    \includegraphics[scale=0.3]{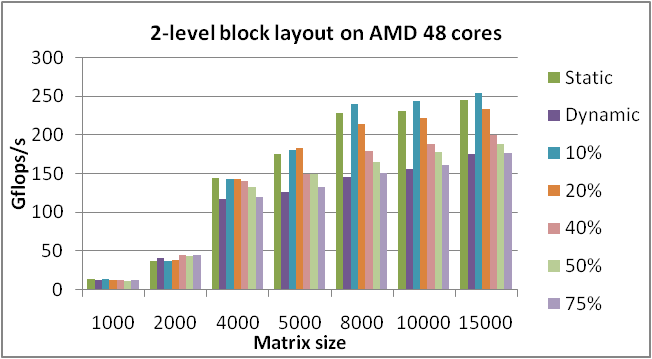}
    \caption{Performance of CALU with static/dynamic scheduling on AMD Opteron 48 core machine. The percentage of the dynamic part is varied from 10\% to 75\%. The matrix is stored using 2-level block layout}
    \label{fig:2lbl_amd_all_48cores}
  \end{center}
\end{figure}

Figure \ref{fig:2lbl_amd_all_48cores} shows the performance obtained
on the 48 core AMD Opteron machine.  In this case, varying the percentage
of the dynamic part in the hybrid static/dynamic scheduling leads to
important differences in performance. CALU dynamic is the least
efficient approach. There are three main reasons for this. First, the
blocks are stored contiguously in memory such that they fit in cache,
but due to the dynamic scheduling, the data might not be reused.
Second, when the matrix size increases along with the number of
blocks, the dequeue overhead of the dynamic scheduler becomes significant.
Third, due to the storage of the matrix, we do not group blocks
together to improve the performance of BLAS operations and reduce
scheduling overhead.  Due to these reasons, increasing the percentage
of the dynamic part in CALU static/dynamic does not lead to better
performance.

\begin{figure}
   \begin{minipage}[c]{.49\linewidth}
      \includegraphics[height=0.65\linewidth,width=\linewidth]{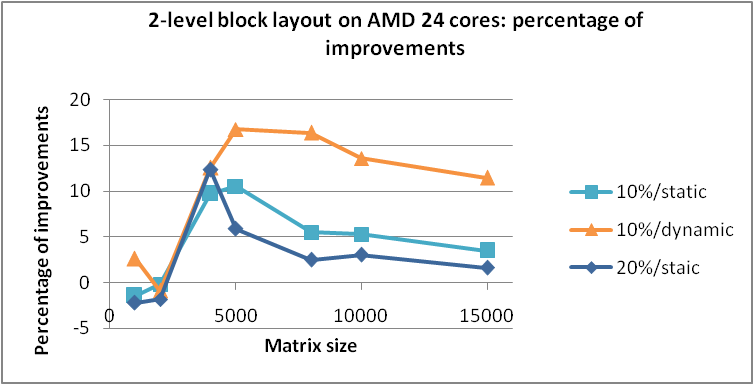} 
      a. Experiments on 24 cores
   \end{minipage} \hfill
   \begin{minipage}[c]{.49\linewidth}
      \includegraphics[height=0.65\linewidth,width=\linewidth]{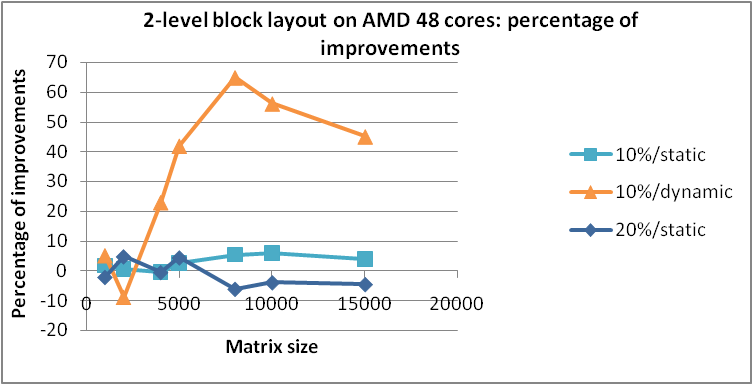}
      b. Experiments on 48 cores
   \end{minipage}
   \caption{Percentage of improvement of CALU static(10\% dynamic) and CALU static(20\% dynamic) over CALU static and CALU dynamic on AMD 48 core machine. The matrix is stored using a 2-level block layout.}
   \label{fig:2lbl_precentage_improvements}
\end{figure}

Figure \ref{fig:2lbl_precentage_improvements} shows the percentage of
improvement of CALU static(10\% dynamic) and CALU static(20\% dynamic)
over CALU static and CALU dynamic. In the best case, CALU static(10\%
dynamic) is 5.9\% faster than static, and 64.9\% faster than dynamic on
48 cores. On 24 cores, CALU static(10\% dynamic) is up to 10\% faster
than CALU static, and up to 16\% faster than CALU dynamic.

\subsubsection{Summary of results}

\begin{figure}[htbp]
  \begin{center}        
    \includegraphics[scale=0.3]{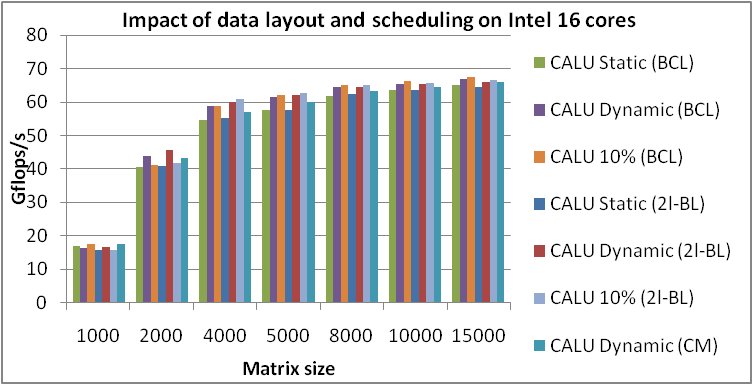}
    \caption{Impact of data layout and scheduling on the Intel 16 core machine.}
    \label{fig:intel_data_layout_impact_16cores}
  \end{center}
\end{figure}

\begin{figure}[htbp]
  \begin{center}        
    \includegraphics[scale=0.3]{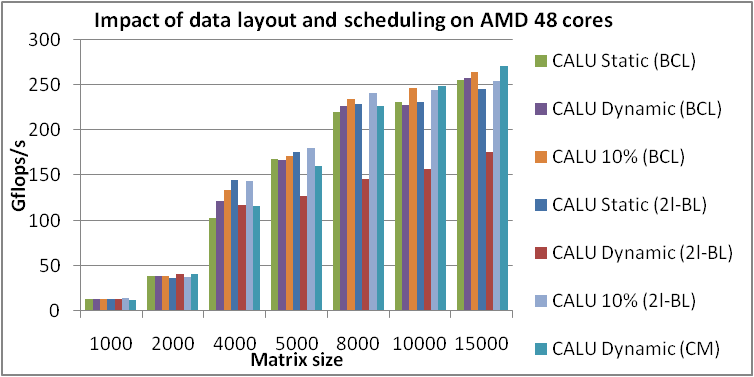}
    \caption{Impact of data layout and scheduling on AMD 48 core
      machine.}
    \label{fig:amd_data_layout_impact_48cores}
  \end{center}
\end{figure}

Figures \ref{fig:intel_data_layout_impact_16cores} and
\ref{fig:amd_data_layout_impact_48cores} show a summary of our results
on both machines.  (In the figures, dynamic rectangular refers to a
column major layout of the input matrix.)  We note that the
performance of the algorithm depends on the matrix size, the data
layout used, and is architecture dependent. However several trends
appear.  On the Intel Xeon machine, the dynamic scheduling is fairly
efficient. For this machine, the time to transfer data from main
memory to the cache of each core does not hinder the performace of a
fully dynamic scheduling strategy. However, on a NUMA machine like the AMD Opteron,
fully dynamic scheduling is highly inefficient due to the cost of cache misses, and
so exploiting locality through constraining data migration is essential on such a
machines.

When the matrix is small ($n \leq 5000$), the two-level block cyclic
layout leads to good performance.  But with increasing matrix size,
the block cyclic layout leads to better performance than the two-level
block layout.  This is mainly due to our approach of performing BLAS 3
operations on larger blocks when the block cyclic layout is used.
When the matrix is large enough, there are enough tasks to schedule,
the synchronization is reduced, and the BLAS 3 operations are more
efficient.

In general, CALU static with a small percentage of dynamic 
leads to the best performance gains, and can achieve performance that is closer to 
peak performance on both machines.
On the Intel machine, CALU static(10\% dynamic) achieves up to
$67.4$ Gflops/s, which is $79$\% of the peak performance.  On the AMD
machine, CALU static(10\% dynamic) achieves up to $264.1$ Gflops/s
that is $49$\% of the peak performance.  Both results were obtained
for a matrix of size $m=n=15000$ stored using a block cyclic layout. 
   
\subsection{Profiling}
\begin{figure}[htbp]
  \begin{center}        
    \includegraphics[angle=0,scale=0.2]{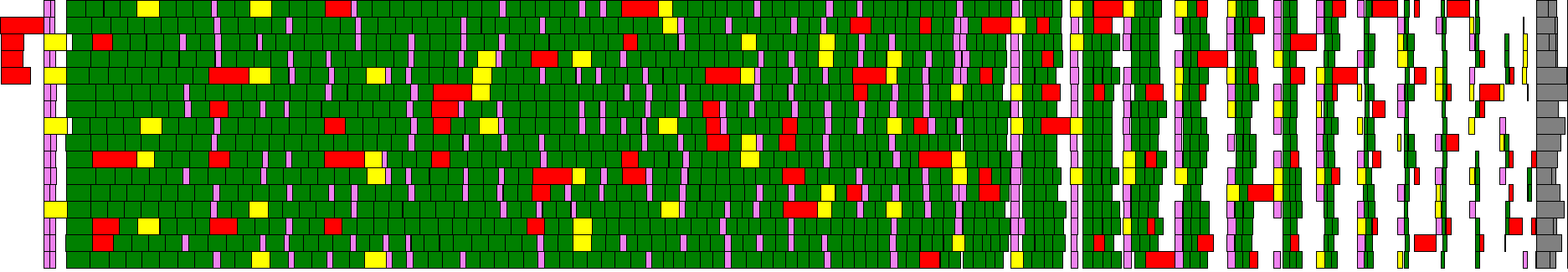}
    \caption{CALU dynamic with column major layout on AMD machine.}
    \label{fig:calu_rectangular_dynamic_16cores}
  \end{center}
\end{figure}

We observe the timelines of our algorithm on a matrix of size $2500
\times 2500$ with a block size of $b=100$ using 16 cores of the AMD
machine. 
Figure \ref{fig:calu_rectangular_dynamic_16cores} shows the profiling
of the dynamic version of CALU using a column major layout. 
We observe that 90\% of threads become idle after only 60\% of the
total factorization time, while for the other variants of scheduling,
this happens towards the very end, after 80\%-90\% of the total
factorization time.

As presented in the Introduction, Figure \ref{fig:calu_static_profile}
shows the profiling of the static version of CALU, we observe pockets
of idle times during the factorization.

\begin{figure}[htbp]
  \begin{center}        
    \includegraphics[angle=0,scale=0.2]{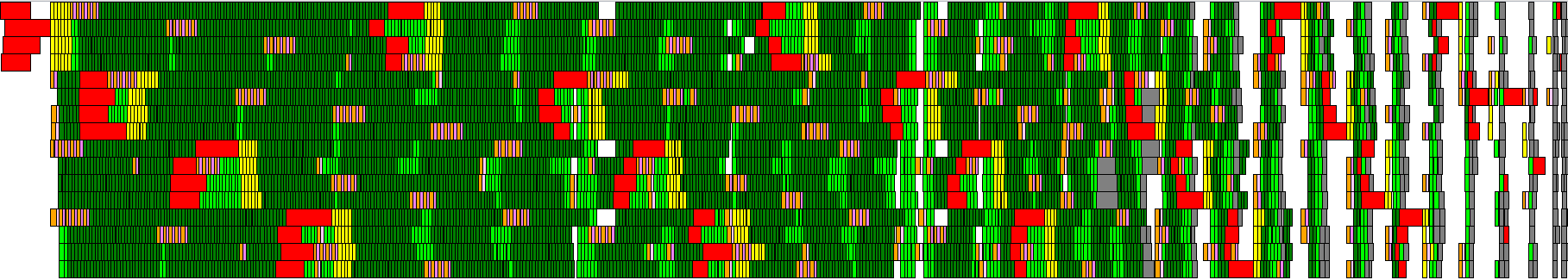}
    \caption{CALU static (10\% dynamic) with 2-level block layout on AMD using 16 cores}
    \label{fig:calu_static_dynamic_10percent}
  \end{center}
\end{figure}

Figure \ref{fig:calu_static_dynamic_10percent} shows the profiling of
static(10\% dynamic) version of CALU.  We observe that a small
percentage of dynamic helps to keep the cores busy, and reduces
drastically the idle time.

\subsection{Comparison with MKL and PLASMA}

\begin{figure}[htbp]
  \begin{center}        
    \includegraphics[scale=0.2]{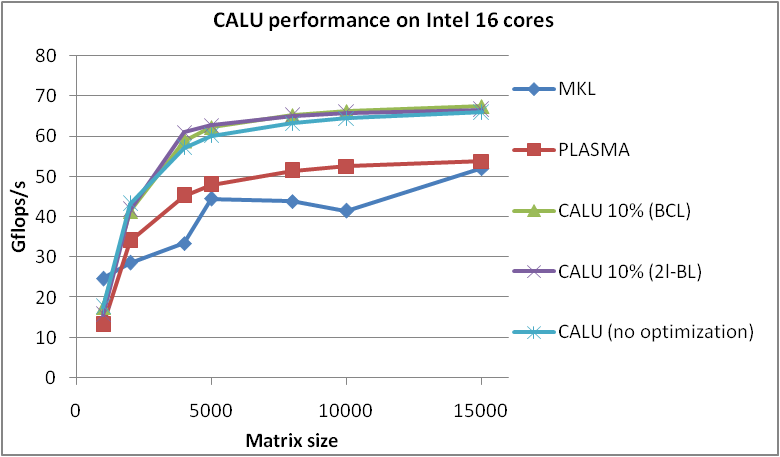}
    \caption{Performance of CALU, MKL, and PLASMA on the 16 core Intel
      machine.}
    \label{fig:intel_calu_performance_16cores}
  \end{center}
\end{figure}

\begin{figure}[htbp]
  \begin{center}        
    \includegraphics[scale=0.2]{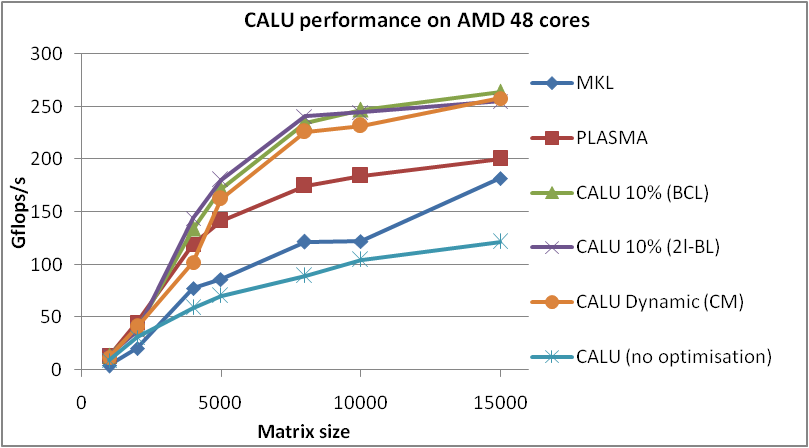}
    \caption{Performance of CALU, MKL, and PLASMA on the 48 core AMD
      Opteron machine.}
    \label{fig:amd_calu_performance_48cores}
  \end{center}
\end{figure}

We compare in Figures \ref{fig:intel_calu_performance_16cores} and
\ref{fig:amd_calu_performance_48cores} the performance of CALU
static(10\% dynamic) against the dgetrf routine from MKL and
dgetrf\_incpiv from PLASMA. The routine from MKL implements Gaussian
Elimination with partial pivoting.  Since the initial data placement
may have a dramatic impact on the performance of an application
running on NUMA machines~\cite{kleen2005numa}, we distribute the input
matrix to all the cores before calling MKL. This distribution was
done using existing numactl (with argument --interleave), which
controls NUMA policy for shared memory. This improves dramatically
the performance of the routine MKL\_dgetrf on the AMD 48 core machine.
 
Our algorithm outperforms MKL on both the Intel and AMD machines.  On the
16 core Intel Xeon machine, for $M=N=10000$, CALU static(10\%
dynamic) with both BCL layout and 2l-BL layout is about $60\%$ faster
than MKL.  The best improvement is obtained with CALU static(10\%
dynamic) (2l-BL) for $M=N=4000$, where it is $82\%$ faster than MKL.
On the 48 core AMD machine, for $M=N=10000$, CALU static(10\%
dynamic) with both data layouts is about $100\%$ (up to $110\%$ faster than
MKL.

We also observe improvements (up to $20\%$ - $30\%$ for larger
matrices) with respect to the dgetrf\_incpiv routine from PLASMA.
This routine implements the LU factorization using incremental
pivoting (which can be seen as a block version of pairwise pivoting,
whose stability is still under investigation), in which the panel
factorization is removed from the critical path.  This leads to a
task dependency graph that is \textit{different}
 from CALU's task dependency graph.
In the recently released version of PLASMA, there is also an implementation of Gaussian
elimination with partial pivoting, but we do not have a thorough
comparison against this new routine for the moment.

\section{Theoretical Analysis}
To understand these results, we provide a basic
performance model and theoretical analysis. 
Because we want to avoid load balancing as much as possible until it
is absolutely needed(due to the overheads it
incurs, such as coherence cache misses and dequeue overhead),
we aim to minimize the percent dynamic. 
Thus, we ask the following question: given a particular algorithm and a
particular architecture, what is the minimum percentage 
dynamic $d_{ratio}$(defined in the algorithm section) that
should be used in an algorithm to obtain the best performance? 
 To understand this, we formalize the problem as follows. Let $f_s$ the fraction of work
done statically. Note the $d_{ratio} = 1-f_{s}$. Let $p$ be the number
of cores. Let $T_{1}$ be the serial time for the computation. Let
$T_{p}$ = $\frac{T_{1}}{p}$ be the time for computation to be done in
parallel across $p$ cores.  
We formalize our question by asking the following: what is the largest
static fraction $f_{s}$ that will make it feasible to attain ideal
execution time $t_{ideal}$, given a compute core $i$ has 
excess work ${\delta}_{i}$\footnote{More realistically, this excess work occurs with
some probability $\phi$, and thus we weight each load imbalance
${\delta}_{i} $ by ${\phi}_{i}$. However, we make the simplifying assumption that we
know that this transient load imbalance will definitely occur. In
other words, our analysis assumes $\phi$ $=$ $1.0$.}? Let ${\delta}_{max}$ be the maximum 
excess work across all cores. Let ${\delta}_{avg}$ be the average excess work across all cores.
Theorem 1 provides the bound for this static fraction.\\ 


\textbf{Theorem 1:}  $f_{s} \leq 1 - \frac{{\delta}_{max} - {\delta}_{avg}}{T_{p}}$ \\
\textit{Proof:}
In the presence of some excess work 
${\delta}_{i}$ (e.g. system noise) that is forced on core $i$,
let $t_{ideal}$ be the ideal time for computation for a given
number of cores $p$ when excess work can be load balanced, and 
let $t_{actual}$ be the worst-case time taken when 
the excess work cannot be load balanced. This means that: \\ 

 $t_{ideal} = \frac{T_1 + \sum_{i=1}^n {\delta_{i}}}{p}$ and $t_{actual} = f_{s} \times \frac{T_1}{p}$ $+$ $\max{ {\delta}_{i}}$ \\  

To find the breakpoint at which static scheduling will induce load imbalance, we set $t_{actual} \leq t_{ideal}$.  
Given this, the time for the case when the compuation is load imbalanced, $t_{actual}$,
will be is no worse than the case of completely load balanced
computation, $t_{ideal}$. Expanding this inequality, we have: \\

$f_{s} \times \frac{T_1}{p}+ \max_{i=1}^p {{\delta}_{i}}$  $\leq$ $\frac{T_1 + \sum_{i=1}^n {\delta_{i}}}{p}$ \\

Solving for the static fraction $f_{s}$, we have: \\

$f_{s} \leq  \left(\frac{T_1 + \sum_{i=1}^n {\delta_{i}}}{p} - \max_{i=1}^p {\delta_{i}} \right) \times \frac{p}{T_1}$  \\ 

$=$ $f_{s} \leq \frac{T_1 + \sum_{i=1}^n {\delta_{i}}}{T_1} -
  \frac{\max_{i=1}^p {\delta}_{i} \times p}{T_1}$  \\ 

$=$ $f_{s} \leq  1  -  \frac{ { \left( \max_{i=1}^p {\delta}_{i} \right) \times p} -  \sum_{i=1}^n {\delta_{i}} }{T_1} $ \\

$=$ $f_{s} \leq 1  -  \frac{ { \left( \max_{i=1}^p {\delta}_{i} \right)} - \frac{\sum_{i=1}^n {\delta_{i}}}{p}}{\frac{T_1}{p}} $ \\

Based on our assumptions of $T_p$, and our definition of  $\delta_{max}$, $\delta_{avg}$:  \\

 $f_{s} \leq 1 - \frac{\delta_{max} - \delta_{avg}}{T_{p}}$ 

Note that this analysis assumes that parallel time includes no
overheads. Due to the communication on critical path of LU(even for
the communication-avoiding case), a full analysis of our LU factorization
cannot ignore the term of communication cost, $T_{criticalPath}$. 
If  $p \prec $ $\frac{T_{1}}{T_{criticalPath}}$, the $T_{criticalPath}$ does 
not dominate the total execution time of parallel CALU.
Our analysis presented is easily extensible though for the case 
when $p$ $\geq$ $\frac{T_{1}}{T_{criticalPath}}$; this 
term of communication cost can be added to the denominator. 
Thus, the denominator will be $\frac{T_{1}}{p} + T_{criticalPath}$. 
If we also assume there is a cost of migration of tasks
$T_{migration}$(due to coherence cache misses that scheduling incurs),
then the denominator becomes $\frac{T_{1}}{p} + T_{criticalPath} +
T_{migration}$.  The model can be made even more accurate by
incorporating other costs of load balancing (e.g. dequeue overheads)
as well. In general, these additional relevant costs can be captured
by adding a single term, $T_{overhead}$ to the term in the denominator
$\frac{T_{1}}{p}$, effectively providing a more realistic value for
$T_{p}$ that incorporates both communication cost and load balancing
cost.

This simple theoretical analysis 
allows us to more clearly understand the impact of 
application parameters to our experimental results.
Given the time complexity of the parallelized version of the computation
$T_{p}$ of a dense factorization (again assuming there is no cost of 
data movement),the above formula gives us the ability to
plug in the expression for $T_p$ to find the upper-bound 
on the static fraction.
Increasing matrix size can cause an increase in $T_1$ in
Theorem 1. From Theorem 1, we see that 
increasing matrix size allows us to increase the 
maximum static fraction that we can use.
In general, as the total cost of the algorithm  $T_{1}$ increases and we keep two architectural parameters 
$p$ and $\delta_{max} -\delta_{avg}$ constant, we can use a larger percent static 
to avoid scheduling overheads.  

The static fraction can also be affected by architectural parameters.
On the Intel machine, for example, communication 
compared to computation is negligible,
due to the low-latency of coherence cache miss. 
This decreases percentage of dynamic fraction, and increases the
static fraction $f_s$. Thus, as the penalty for coherence cache 
misses becomes higher, the percentage static will need to increase to avoid such coherence
cache misses.
\section{Discussion}



A more detailed performance model and theoretical analysis for regular mesh codes 
has been established through the follow-up studies of the work by 
V. Kale et al \cite{vivek2010load}. 
The adoption of this model will allow us to  calculate expected
completion time of a dense matrix factorization in the 
presence of transient load imbalance occurring with some probability. 
Given this, we can couple our performance model with
auto-tuning techniques and heuristics\cite{LBM-tuning}
for optimizing the scheduling strategy for a particular architecture,
allowing us to significantly prune the search space of parameter
configurations for our scheduling approach.





The problem of noise amplification has been projected (most recently
through simulation) to seriously impede performance at very large scale~\cite{hoefler-noise-sim}. 
In earlier work~\cite{vivek2010load}, it was shown that by improving performance consistency
of the 3D regular mesh code, one can mitigate the impact
of the noise amplification problem.  The early results of the reduced standard deviations of wall
clock times across multiple runs of our code under our tuned
scheduling strategy is in accord with the performance consistency results shown in ~\cite{vivek2010load}. 

With our theoretical analysis, we can provide projections for the
upper-bound on the static fraction to be used within many-core nodes
of an exascale machine.  Keeping the work per core constant, the
term
$ \left( \delta_{max} -\delta_{avg} \right)$ can increase in the
presence of noise amplification.  
Given this and using Theorem 1, we project that the lower-bounds for
percentage dynamic for numerical linear algebra routines will have to
increase for use on future high-performance clusters.



\section{Related work} 
The idea of combining static and dynamic scheduling has been studied
in different contexts in the literature, but to the best of our
knowledge none of the approaches focuses on dense numerical linear algebra.
V. Kale et al. suggested a hybrid static/dynamic approach in
\cite{vivek2010load} that can be incorporated into current MPI
implementations of regular mesh codes to improve the load balancing of
the initial static decompositions. They show the performance of their
strategy on a 3D jacobi relaxation problem.  Our work here embraces
the fundamental principles advocated in that paper, and applies it in
the context of dense matrix factorizations.
Xue et al. introduced an approach in \cite{xue2007locality} that
improves the data locality when executing loop iterations in
codes. This is done in the context of chip multi-processors.  The
authors show that the different loops of many codes may be decomposed
into two parts: in one part, iterations are distributed across
processors at compilation time; in the other part, iterations are
distributed at runtime to available processors to improve the load
balancing.  

Our approach has some similarity with work-stealing, but proceeds more
efficiently.  In work stealing, the work is initially (statically)
distributed almost equally to each thread. During the algorithm, each
inactive thread (the thief) can pick a task from the queue of tasks of
another thread (the victim).  An important question is: from which
other thread will this thread steal work? The approach of randomized
work-stealing from the queue of another thread is implemented in
Cilk~\cite{blumofe1995cilk}. Cilk is founded on theoretical analysis
and proofs on its efficiency.  It has been shown to offer acceptable
performance, particularly for multi-programmed workloads on
multi-cores. 

In the LU factorization, the update of the trailing sub-matrix is
performed from left to right, to maintain the execution of the
algorithm on the critical path and to guarantee that the panel
factorization will be able to start as soon as possible even if the
updates of the columns of the previous steps are not completely done.
In other words, columns close to the panel have high priority.
Although random stealing can help balancing work among processors
during execution, it might not follow the critical path of the
factorization.
In typical implementations of work stealing, when the queue of a
particular thread is empty, that thread attempts to pick a task either
at the top, or at the bottom, of a victim queue. Whether chosen from
the top or the bottom of the queue, this approach is not optimal for
many computations and in particular for dense factorizations. Picking
a task at the beginning of the queue (FIFO) may lead to non-negligible
synchronization overheads with the victim, or may cause false sharing
due to two threads accessing data in two regions in memory in close
proximity.  While picking a task at the end of the queue (LIFO) may
lead to acceptable load balancing in general, for computations in
LU/CALU, the last columns have the least priority within the
computation. This inhibits the progression of the critical path. 
Other work such as~\cite{rudolph1991simple} suggest
then the steal is done from the queue of the
thread having the highest number of tasks.
This cannot be applied directly to dense factorizations, 
where the number of tasks of a thread is not proportional to its workload.

\section{Conclusions}

We have designed and implemented a strategy that combines static and
dynamic scheduling to improve data locality, load balancing,
and exploit the power of current and emerging multi-core
architectures.  Our hybrid static/dynamic scheduling strategy applied
to CALU leads to performance improvements over the fully static CALU
and fully dynamic CALU, and also provides performance improvements
over the corresponding routines in MKL and PLASMA.  
Our performance results show that the combination of static and
dynamic scheduling is effective for dense communication avoiding LU
factorization. In our experiements, we determine the best percentage
of the dynamic part by running variations of the algorithm with
different dynamic percentages. We show that, usually, 10\% dynamic
leads to good performance because it provides the best compromise
between data locality, load balancing, and minimal dequeue overhead. 
While in this paper we focus on CALU, the same techniques can be applied to
other dense factorizations as Cholesky, QR, rank revealing QR, $L D
L^T$, and their communication avoiding versions.  This remains future
work.


On future high-performance clusters planned within the timespan of 5
years, often termed as exascale, each node will be comprised of many
levels of parallelism, with possibly on the order of 100s of cores per
node.  Choosing between purely static and purely dynamic scheduling
for numerical linear algebra routines will escape the problem of
trying to gradual evolve, rather than radically change, our codes for 
use on emerging and future high-performance clusters.
 Our hybrid approach allows for self-adaptivity of the numerical linear
algebra routines to the transient dynamic variations of the
architecture, \textit{without} loss of data locality that is so
fundamentally important to a large class of HPC applications. 

In future work, we opt to design a stronger model to determine the
percentage of the dynamic section. We provide an approach to
theoretically determine the percentage of the dynamic section, but we
believe we can obtain even tighter bounds, given knowledge of
scheduling costs. Furthermore, this theoretical analysis can be
applied to other numerical linear algebra routines, as well as to
full-fledged applications, given the expression for parallel execution
time. We plan to enhance our scheduling technique so that tasks are
chosen from the queue such that the data that these tasks operate on
is highly likely to be in a core's cache already, allowing for fewer
coherence cache misses due to task migration.  Such intelligent
selection of tasks can be implemented by adding locality tags within
the task data structure, and incorporating heuristics that make more
accurate predictions on the task that is least likely to incur a
migration overhead, if chosen.

\section*{Acknowledgment}
We would like to thank James Demmel for his early input and support for the importance of data locality to the scheduling scheme. 
We would also like to thank Jack Dongarra for the access to the machines at Tennessee.
 This work was done in the context of the INRIA-Illinois joint-laboratory of PetaScale Computing.

\nocite{*}
\bibliographystyle{plain}
\bibliography{pf_bib}

\end{document}